\documentclass[conference]{IEEEtran}
\IEEEoverridecommandlockouts
\usepackage{cite}
\usepackage{amsmath,amssymb,amsfonts}
\usepackage{algorithmic}
\usepackage{graphicx}
\usepackage{textcomp}
\usepackage{xcolor}
\usepackage{multirow}
\usepackage{booktabs}
\usepackage{listings}

\def\BibTeX{{\rm B\kern-.05em{\sc i\kern-.025em b}\kern-.08em
    T\kern-.1667em\lower.7ex\hbox{E}\kern-.125emX}}
\begin{document}
\onecolumn
\begin{flushleft} 
“© 2019 IEEE.  Personal use of this material is permitted. Permission from IEEE must be obtained for all other uses, in any current or future media, including reprinting/republishing this material for advertising or promotional purposes, creating new collective works, for resale or redistribution to servers or lists, or reuse of any copyrighted component of this work in other works.”
\end{flushleft}.
\twocolumn

\IEEEoverridecommandlockouts
\IEEEpubid{\makebox[\columnwidth]{978-1-7281-2769-9/19/\$31.00 ~\copyright2019 IEEE \hfill} \hspace{\columnsep}\makebox[\columnwidth]{ }}

\title{Accelerating Transient Fault Injection Campaigns by using Dynamic HDL Slicing}

\author{\IEEEauthorblockN{Ahmet Cagri Bagbaba\IEEEauthorrefmark{1}\IEEEauthorrefmark{2},
Maksim Jenihhin\IEEEauthorrefmark{2}, Jaan Raik\IEEEauthorrefmark{2}, Christian Sauer\IEEEauthorrefmark{1}}
\IEEEauthorblockA{\IEEEauthorrefmark{1}Cadence Design Systems,
Munich, Germany; \IEEEauthorrefmark{2} Tallinn University of Technology, Tallinn, Estonia \\
Email: \IEEEauthorrefmark{1}\{abagbaba, sauerc\}@cadence.com,
\IEEEauthorrefmark{2}\{maksim.jenihhin, jaan.raik\}@taltech.ee}}
\maketitle

\begin{abstract}
Along with the complexity of electronic systems for safety-critical applications, the cost of safety mechanisms evaluation by fault injection simulation is rapidly going up. To reduce these efforts, we propose a fault injection methodology where Hardware Description Language (HDL) code slicing is exploited to accelerate transient fault injection campaigns by pruning fault lists and reducing the number of the injections. In particular, the dynamic HDL slicing technique provides for a critical fault list and allows avoiding injections at non-critical time-steps. Experimental results on an industrial core show that the proposed methodology can successfully reduce the number of injections by up to 10 percent and speed-up the fault injection campaigns.
\end{abstract}


 \IEEEpubidadjcol

\section{Introduction}

With new and increased capabilities in applications such as autonomous driving, the complexity of electronics systems for safety critical applications is growing exponentially. This is causing a shift in the traditional design flow and is pushing ISO26262 compliance down in the semiconductor chain to the individual IP provider and even into the traditional Electronic Design Automation tools. As a result, functional safety compliance becomes a part of the requirements for the development of complex electronics systems. During the design of ISO26262 compliant chips, designers need to evaluate effectiveness of the design to deal with random hardware failures. This is usually done by Fault Injection Simulations. Also, ISO26262 standard highly recommends using of fault injection during the development process of integrated circuits~\cite{ISO2}.

Fault injection is a powerful technique that shows the behaviour of a circuit under the effect of a fault~\cite{364536}. The objective of fault injection is to mimic the effects of faults originating inside a chip as well as those affecting external buses. Different approaches to fault injection and dependability evaluation have been proposed. These include emulation-based fault injection using FPGA architectures as hardware accelerators to speed up estimation of systems' fault tolerance~\cite{Civera2002},~\cite{4751886} and formal method based approaches~\cite{1498171},~\cite{4479838}. This paper focuses on the simulation-based fault injection approach, which can be applied to larger designs compared to the formal and emulation-based solutions.

Having enormous number of possible faults in modern designs is a major drawback of simulation-based fault injection technique as designers need to execute a fault-free simulation as well as thousands of faulty simulations~\cite{998398}. Therefore, it is too hard to inject all possible faults in an acceptable time in all possible locations and at each clock cycle~\cite{5090716}. One solution is to use Statistical Fault Injection (SFI)~\cite{5090716} in which only a randomly selected subset of possible faults is injected. SFI can provide a better execution time by reducing the number of the injections with an error margin. Moreover,~\cite{newJ} have demonstrated that with randomly selected fault lists the ratio of faults which do not produce errors may range as low as 2 to 8 percent, depending on the design under simulation. In consequence, minimization of fault injection locations or pruning fault lists are  advantageous ways to reduce the fault injection simulation time significantly while allowing injection of a considerably larger number of relevant faults.

This work proposes a simulation-based fault injection methodology based on Dynamic HDL Slicing to minimize the number of fault injections. The proposed methodology identifies critical faults which cause the system to fail in the absence of a safety mechanism, and injects only critical faults during the transient fault injection simulation campaigns. Using critical faults to estimate fault coverage eliminates the possibility of fault injection experiments to produce no error. The main contribution of this work is three-fold as follows:

\begin{figure*}[h]
\centerline{\includegraphics[scale=1.0]{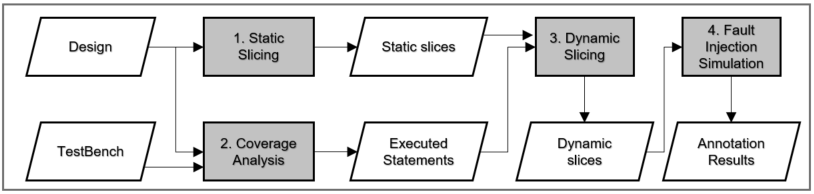}}
\caption{Proposed HDL slicing based fault injection methodology~\cite{8474167}.}
\label{flow}
\end{figure*}

\begin{itemize}
  \item Dynamic slicing on HDL to generate critical fault list
  \item Implicit fault collapsing within the slicing model: The fault list obtained by the proposed slicing method has an additional feature of avoiding injections at time-steps as data inside registers is not being consumed.
  \item Language-agnostic RTL fault injection supported by industrial grade EDA tool flow
\end{itemize}

As a result, this method can successfully reduce the number of fault injections on an industrial core. The fault model implemented in this paper is based on single-clock-cycle bit-flip faults within the RTL registers. This fault model is targeting single Single-Event-Upsets (SEUs) in all the registers of the design. The proposed methodology is demonstrated on Cadence tools but it remains applicable to other tool flows as well. This work is an extension of our previous work~\cite{8474167}. The major difference is that we extend dynamic slices' scope by including both sequential and combinational parts as it is explained in Section~\ref{implicitsection}. This brings 100\% accuracy in the results. In the previous work, less accuracy is adopted as only sequential parts are considered in dynamic slices. The second difference is that we evaluate our methodology in the industrial size CPU with different workloads to demonstrate the potential of the proposed methodology.


This paper is structured as following. Section~\ref{relatedworks} gives an overview of related works. We describe our dynamic HDL slicing methodology in Section~\ref{tecnique}. Experimental results are shown in Section~\ref{ExperimentalResults}. Section~\ref{conclusions} concludes this paper.

\section{Related Works}\label{relatedworks}

There exist many advanced tools and methods for simulation-based fault injection. In~\cite{614074}, a tool called VERIFY (VHDL-based Evaluation of Reliability by Injection Faults Efficiently) is presented that utilizes an extension of VHDL for describing faults correlated to a component, enabling hardware manufacturers, which provide the design libraries, to express their knowledge of the fault behaviour of their components. Although it provides multi-threaded fault injection as well as checkpoints and comparison with a golden run to speed up the simulation of faulty runs, the drawback is that it requires modification of the VHDL language itself. \cite{689479} proposes MEFISTO-C: A VHDL-based fault injection tool that conducts fault injection experiments using VHDL simulation models. A variety of predefined fault models are supported by the tool; however, it does not provide specific optimizations to speed up the simulation.

Several approaches to generate the critical fault list to be considered as the basis of fault list injection have been proposed. In~\cite{513279}, a method for generating a critical fault list is presented. The system under test is described by a data flow graph, the fault tree is constructed by applying the instruction set architecture fault model to the data flow description with a reverse implication technique, the fault injection is performed, and fault collapsing on the fault tree is employed. The proposed method is very costly in terms of CPU time and it therefore not applicable to systems with high complexity.

\cite{998398} presents a new technique and a platform for accelerating and speeding-up simulation-based fault injection in VHDL descriptions. Use check-pointing to reload the fault-free state if the design allowing to start the fault simulation from the clock-cycle of fault injection. In addition, a golden-run fault collapsing technique is utilized that discards all fault injections between read-write and write-write operations of the memory elements. However, the approach does not take advantage of the dynamic slicing benefits.
\cite{newJ} proposes fault collapsing based on extracting high-level decision diagrams from the VHDL model. Although significant speed-up can be achieved, the step of efficient decision diagram synthesis from the full synthesizable subset of VHDL remains an issue.

There are several papers dealing with transient fault injection.~\cite{7684076} shows the results collected in a series of fault injection experiments conducted on a commercial processor. Here, the authors inject a fault in a given sequential element at a given instant of time. However, as it is hard to inject a fault in each of the tens of thousands sequential elements in the processor, the execution is divided into the parts and, for each of these parts, a random fault injection instant is selected.~\cite{8347235} analyses fault injection campaign in the CPU registers by choosing a random instant when the fault is injected.~\cite{6523691} identifies the optimal set of flip-flops but injection time is randomized uniformly over the active region of the simulation. Similarly,~\cite{6734986} injects a fault randomly in time and location in RT-level. Lastly,~\cite{7906763} deals with single and multiple errors in processors by randomly selecting injecting time and choosing registers. As opposed to these works, our approach shows the fault injection time explicitly instead of random instants.

Dynamic slicing technique is used in~\cite{6549113},~\cite{6233020}. The former uses dynamic slicing for statistical bug localization in RTL. The latter proposes dynamic slicing and location-ranking-based method for accurately pinpointing the error locations combined with a dedicated set of mutation operators.

Different from the works listed above, this paper proposes a dynamic HDL slicing based technique that implicitly covers the golden run fault collapsing, thereby significantly speeding up the fault injection process.

\begin{figure*}[h]
\centerline{\includegraphics[scale=1.0]{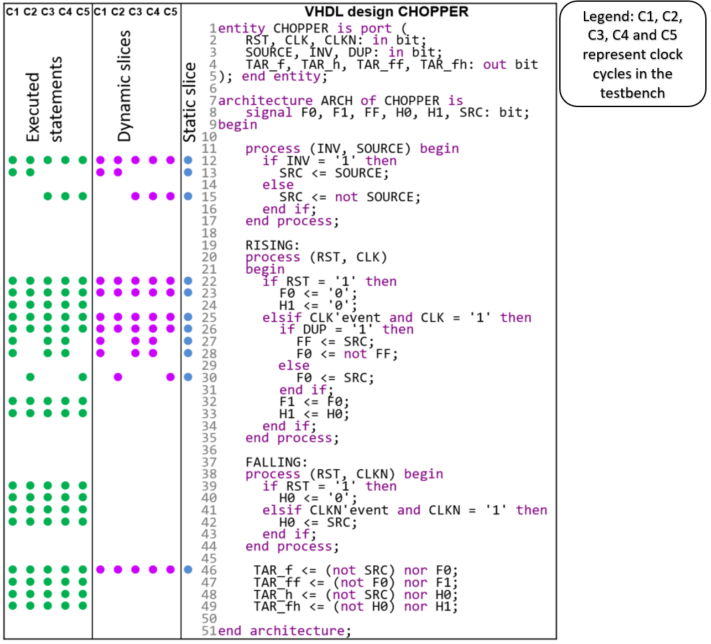}}
\caption{HDL slicing on a motivational example chopper~\cite{8474167}.}
\label{mot}
\end{figure*}

\section{Fault Injection based on Dynamic HDL Slicing Technique}\label{tecnique}

In this work, fault injection simulation campaigns are optimized by pruning the fault list to the critical faults identified using HDL slicing on the RTL design model. The proposed flow is shown in Fig.~\ref{flow} and starts with the (1) extraction of static slices for the target observation point. In parallel, code coverage data is generated by (2) simulation-based code coverage analysis for the design with pre-defined stimuli in the testbench. Next, (3) the dynamic slicing procedure identifies the intersection of the identified static slice and covered code items and results in a set of clock-cycle-long dynamic slices for the given observation point. Finally, (4) the fault injection simulation selects critical faults from the dynamic slices, injects them at the specified time and evaluates the fault propagation. We explain the details of the methodology in the following subsections using a motivational example depicted in Fig.~\ref{mot}, i.e. a VHDL implementation of a signal \emph{chopper} design~\cite{Clarke2002ProgramSF}. Following subsections explain each step of the proposed methodology in detail.

\subsection{Static Slicing}

\begin{figure}[b]
\centerline{\includegraphics[scale=0.6]{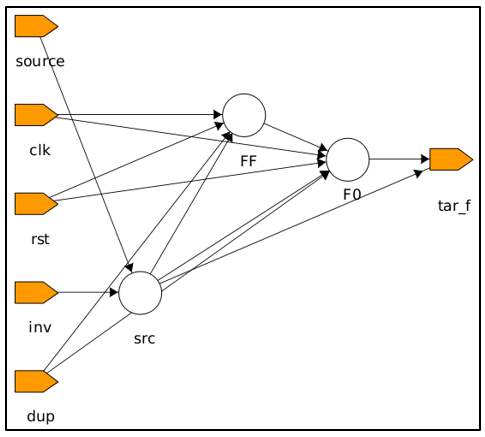}}
\caption{Backward static slice on the signal \emph{TAR\_F} in the chopper design.}
\label{ss}
\end{figure}

Static slice, as it is implemented in the current paper, includes all statements that affect the value of a variable \emph{v} for all possible inputs at the point of interest, e.g., at the statement \emph{x}, in the program. In the RTL code, static slice shows the dependency between HDL statements~\cite{Iwaihara96programslicing}. A simple design \emph{chopper} in Fig.~\ref{mot} has four outputs representing different chops for the input signal \emph{SOURCE} based on the design configuration by inputs \emph{INV} and \emph{DUP}. It is possible to perform a search backward to find dependencies in the HDL. The resulting static slice is computed for the \emph{chopper} design's output \emph{TAR\_F} as shown in Fig.~\ref{ss} by the help of formal analysis tool's structural analysis capability. The column Static Slice in Fig.~\ref{mot} marks HDL statements of a static slice on the \emph{TAR\_F} output. For instance, as the static slice of \emph{TAR\_F} does not include Line 40, \emph{H0} is counted as outside of the static slice and for a \emph{TAR\_F} output there is no need to inject fault on \emph{H0}. Fig.~\ref{mot} also implies that, static slice does not depend on clock cycles (shown as C1, C2, C3, C4 and C5) while executed statements and dynamic slice may change for each clock cycle. In summary, static slice includes statically available information only as it does not make any assumptions on inputs. Static slice is the first step of the proposed methodology to prune fault list.

\subsection{Coverage Analysis}
In parallel to static slicing step, the RTL design is simulated in the logic simulation tool to dump and analyse the coverage data. In this step, we dump coverage data for each clock cycle so that we can find what statements in the RTL are executed for each clock cycle. In the proposed methodology, one clock cycle defines the size of our dynamic slice. We use coverage tool and coverage metrics in order to find executed statements. After loading a simulation run into the coverage tool, we can analyze coverage metrics data scored in that run. In this work, we use code coverage which measures how thoroughly a testbench exercises the lines of HDL code. Code coverage includes block coverage, branch coverage, statement coverage, expression coverage, and toggle coverage. All these coverage types except toggle coverage can be used in this work. Block coverage identifies the lines of code that get executed during a simulation run. It helps us determine if the testbench executes the statements in a block. Branch coverage complements block coverage by providing more precise coverage results for reporting coverage numbers for various branches individually. Statement coverage is just a subset of block coverage and it shows execution of all the executable statements in the RTL. Expression coverage provides information on why a conditional piece of code was executed. At the end of this step, we generate executed statements data to find dynamic slices in the next step. Fig.~\ref{mot} shows executed statements for five clock cycles (C1, C2, C3, C4, C5).

\subsection{Dynamic Slicing} \label{dynamic_explanation}

Dynamic slice, as it is implemented in the current paper, includes those statements that actually affect the value of a variable \emph{v} for a particular set of inputs of the RTL so it is computed on a given input~\cite{Korel:1988:DPS:56378.56386}. It provides more narrow slices than static slice and consists of only the statements that affect the value of a variable for a given input.

In a nutshell, dynamic slice is the intersection of static slice and executed statements. We illustrate the concept of dynamic slice in Fig.~\ref{mot}. This figure also shows how dynamic slices narrow down the fault space when compared to state-of-the-art static slice approach. For instance, during the time window C5, register \emph{FF} (Line 27) is not in dynamic slice meaning that we do not need to inject fault in \emph{FF} at C5 time window. Dynamic slice gives us critical faults and eliminates those faults that are not critical. In this way, we manage to reduce fault list by injecting only critical faults. This provides a speed-up in the fault injection simulation time as each injected fault increases total run time of fault injection campaign.

\subsubsection{Implicit Fault Collapsing in Dynamic Slices} \label{implicitsection}
In our proposed methodology, dynamic slices cover both sequential and combinational parts. In this way, all faults outside of dynamic slices are 100\% undetected and can be collapsed to exclude them from the fault list. When considering the average CPU time per a fault, an undetected fault spends more CPU time than a detected fault as the fault injection simulation for an undetected fault lasts until the end of the simulation. Hence, it is very effective to identify undetected faults without running fault injection campaigns.

\begin{figure}[b]
\centerline{\includegraphics[scale=0.7]{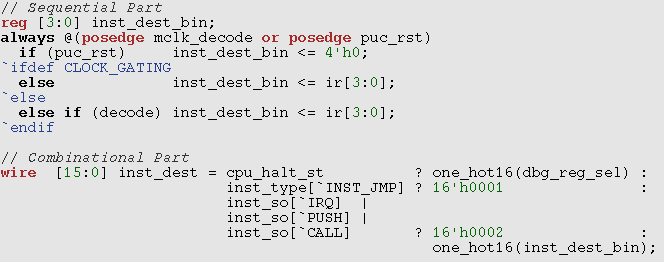}}
\caption{Implicit fault collapsing.}
\label{implicit}
\end{figure}

In the previous work~\cite{8474167}, only sequential parts are considered in dynamic slices; however, both registers (sequential) and combinational parts that are connected to the registers are counted in dynamic slices in this work. In Fig.~\ref{implicit}, dynamic slice is built by considering the register \emph{inst\_dest\_bin} and \emph{inst\_dest} (combinational) so that we can have 100\% accurate results. This is called as implicit fault collapsing since we avoid injections at time-steps as data inside registers is not being consumed.

\subsection{Fault Injection Simulation}

Fault injection enables to verify the capability of a safety mechanism to recognize failures in a design's functionality, by injecting faults into the design. In a fault injection simulation, target system and the possible hardware faults are modeled and simulated by the simulator. In this process, the system behaves as if there is a hardware fault.

To inject faults into a design, fault injection simulator needs to know fault target at which to inject fault. In this work, we enable fault instrumentation on the dynamic slices, more specifically on registers that are in dynamic slices. In other words, the proposed method identifies critical faults from dynamic slices and inject them at the specified times. As a fault model, we use Single Event Upset (SEU) fault type which inverts the value of output of a sequential element and hold the modified value until it is assigned a new value. Another thing that fault injection simulations need is an \emph{observation point}, since the purpose of a fault campaign is to verify that an error will be observed at some specific point in the design. By defining explicit observation points when running a good simulation, we can generate data that will help us to determine if an injected fault is detected or undetected at one or more specified nets.

In brief, fault injection simulation is used to show the effectiveness of the proposed method. We inject one fault in one simulation run. Also, in the case of having more than one observation point in the analysis, the proposed method prevents multiple injection of faults within the overlap of static slices.

\section{Experimental Results} \label{ExperimentalResults}

In order to verify the accuracy of the proposed fault injection method, we evaluate our application on industrial CPU with different workloads~\cite{opencores}. In the following subsections, firstly, we explain our experimental setup. Then, we show the results in detail.

\begin{figure}[b]
\centerline{\includegraphics[scale=0.45]{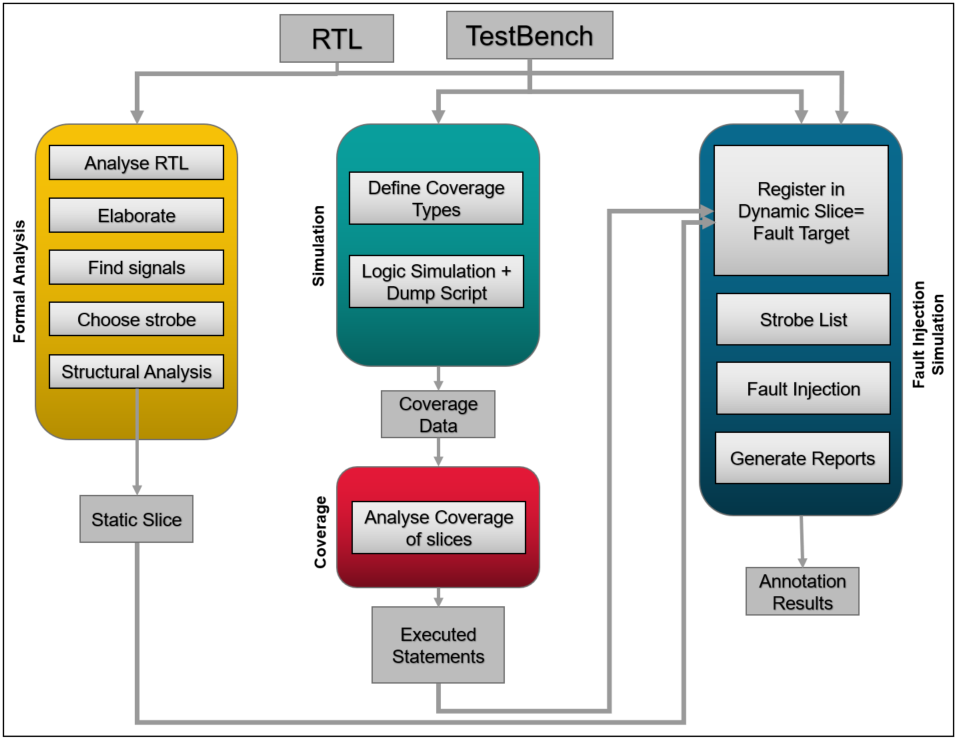}}
\caption{Overall flow of experimental setup.}
\label{fl}
\end{figure}

\subsection{Experimental Setup}

Aiming to automate the execution of fault injection campaigns using the different tools, an application is developed as in the Fig.~\ref{fl}. This is the more detailed illustration of Fig.~\ref{flow}. We create generic scripts to activate the tools and automate the flow. In this work, the proposed methodology is integrated into Cadence flow but it can be applied using tools by any major EDA vendor.

In the first step, backward static slice is built for a selected observation point by using Cadence{\textregistered} JasperGold Formal Verification Platform. Then, we generate coverage results through Cadence{\textregistered} Xcelium{\texttrademark} for each clock cycle that defines the size of the dynamic slices. In the next step, static slice and executed statements data are sent to fault injection simulation to define fault target for the campaign. Annotation results provide information regarding to number of injected faults, number of detected and undetected faults. Moreover, we also use the profiling feature of the tool that measures where CPU time is spent during simulation. The profiler generates a run-time profile file that contains simulation run-time information that is useful for comparing execution time of different campaigns. Cadence{\textregistered} Xcelium Fault Simulator is used for fault injection simulations.

\newcommand{\ra}[1]{\renewcommand{\arraystretch}{#1}}
\begin{table*}[h]
\centering
\ra{1.3}
\caption{Experimental Results on openMSP430}
\label{tab:allResults}
\resizebox{\textwidth}{!}{%
\begin{tabular}{lcccccccc}\toprule

\multicolumn{1}{c}{} & \multicolumn{2}{c}{Sandbox} &  & \multicolumn{2}{c}{Dhrystone} &  & \multicolumn{2}{c}{Coremark} \\ \cline{2-3} \cline{5-6} \cline{8-9}
\multicolumn{1}{c}{} & Static Slicing & Dynamic Slicing &  & Static Slicing & Dynamic Slicing &  & Static Slicing & Dynamic Slicing \\ \hline
\multicolumn{9}{l}{$inst\_dest\_bin$} \\
Detected & 8036 & 8036 &  & 56236 & 56236 &  & 48891 & 48891 \\
Undetected & 3996 & 2852 &  & 43764 & 42404 &  & 51109 & 47809 \\
Total & 12032 & 10888 &  & 100000 & 98640 &  & 100000 & 96700 \\
Total CPU time of overall regression & 1197.1s & 994.7s &  & 658919.7s & 622459.0s &  & 3437663.0s & 3323109.9s \\
Fault Coverage & 66.788\% & 73.806\% &  & 56.236\% & 62.735\% &  & 48.891\% & 50.559\% \\
\multicolumn{9}{l}{$inst\_src\_bin$} \\
Detected & 2423 & 2423 &  & 34766 & 34766 &  & 45161 & 45161 \\
Undetected & 9609 & 8413 &  & 65234 & 63498 &  & 54839 & 48051 \\
Total & 12032 & 10836 &  & 100000 & 98264 &  & 100000 & 93212 \\
Total CPU time of overall regression & 1488.2s & 1284.2s &  & 803009.1s & 790300.0s &  & 3575198.1s & 3178378.2s \\
Fault Coverage & 20.137\% & 22.361\% &  & 34.766\% & 35.380\% &  & 45.161\% & 48.450\% \\
\bottomrule
\end{tabular}%
}
\end{table*}

\subsection{Evaluation and Results}

We evaluate our methodology on a 16-bit microcontroller core~\cite{opencores} with a single address space for instructions and data. To show the effectiveness of the proposed method, we use three different workloads on \emph{openMSP430}. We show our results in two categories as fault list reduction and time savings. Then, we evaluate the accuracy of this methodology by comparing our results to a state-of-the-art static slicing optimization method.

In the first step, backward static slice is built from \emph{dmem\_din} observation point which is the main output of the core and then coverage data is calculated. Next, considering the registers in static slice, instruction source and destination registers are selected as fault targets to apply the proposed method since these registers are widely used in fault injection applications as they hold all instructions.

Table~\ref{tab:allResults} shows the comparison of two techniques: $a)$ state-of-the-art static slicing and $b)$ dynamic HDL slicing. We perform a fault injection campaign for each workload and a fault target (and) for each approach. For the execution of Dhrystone and Coremark workloads with static slicing, we select 100k faults after the warm-up phase of the CPU.

\begin{figure}[t]
\centerline{\includegraphics[scale=0.5]{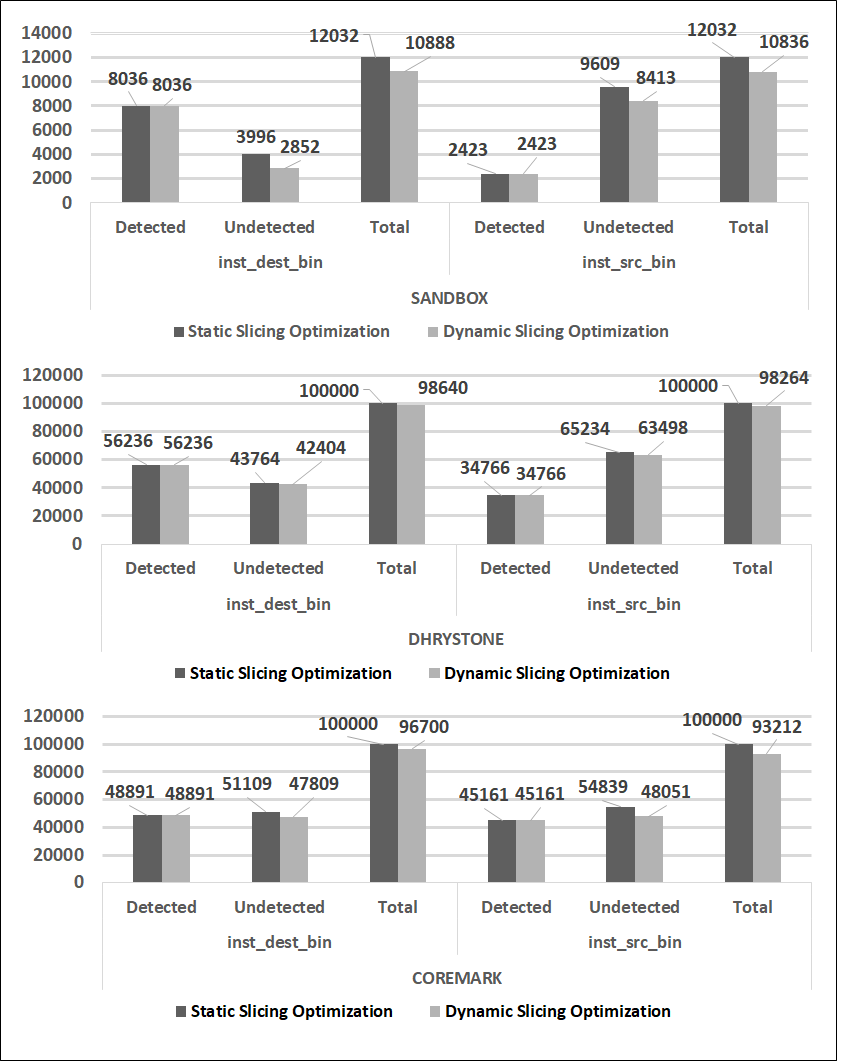}}
\caption{Fault list reduction based on three different workloads.}
\label{graph1}
\end{figure}

\subsubsection{Fault List Reduction}
Fig.~\ref{graph1} shows the reduction in the number of faults injected. All detected faults seen in Fig.~\ref{graph1} are critical faults. As seen in this charts, dynamic HDL slicing is effective in pruning the fault list as compared to the static slicing. Table~\ref{tab:reductions} shows the percent reduction in the number of faults injection. The best reduction is achieved in Sandbox workload as a reduction of 9.94\%. The magnitude of the fault list reduction depends on the workload characteristics. In this experimental results, the fault list reduction varies between 1.36\% and 9.94\%. These analysis reveal that dynamic HDL slicing prune the fault list and identify the critical faults successfully while analysis and optimization effort costs are very minor.
Additionally, to identify undetected faults and exclude them from the fault list provides a increased fault coverage as it can easily be seen in Table~\ref{tab:allResults}.

\subsubsection{Time Savings}
Table~\ref{tab:allResults} shows the total CPU time of overall regression for each fault injection campaign. Dynamic slicing provides various time savings from 1.58\% to 16.91\% as shown in Table~\ref{tab:timesaving}. As in fault list reduction, time savings depend on the workload characteristic. When considering the need of multiple fault injection campaigns in real life applications, this time savings can expeditiously increase.

\subsubsection{Accuracy}
In this work, we show the results of a fault injection campaign performed using dynamic slicing, along with a state-of-the-art static slicing approach. These results reveal that dynamic slicing achieves the same number of detected faults as static slicing campaign. This means that dynamic slicing can be used for different purposes as it is an accurate fault injection methodology. For instance, SFI~\cite{5090716} prunes the fault list in terms of margin of error with a given confidence level. However, dynamic slicing exclude only non-critical faults and find all critical faults with a 100\% accuracy.

\begin{table}[]
\ra{1.3}
\caption{Percentage of reduction of the total number of injections with dynamic HDL slicing}
\label{tab:reductions}
\resizebox{\columnwidth}{!}{%
\begin{tabular}{lclcclccl}\toprule

\multicolumn{1}{c}{\textbf{}} & \multicolumn{2}{c}{Sandbox} &  & \multicolumn{2}{c}{Dhrystone} &  & \multicolumn{2}{c}{Coremark} \\ \cline{2-3} \cline{5-6} \cline{8-9}
\multicolumn{1}{c}{} & \multicolumn{2}{c}{Percentage of reduction} &  & \multicolumn{2}{c}{Percentage of reduction} &  & \multicolumn{2}{c}{Percentage of reduction} \\ \hline
$inst\_dest\_bin$ & \multicolumn{2}{c}{9.51\%} &  & \multicolumn{2}{c}{1.36\%} &  & \multicolumn{2}{c}{3.3\%} \\
$inst\_src\_bin$ & \multicolumn{2}{c}{9.94\%} &  & \multicolumn{2}{c}{1.74\%} &  & \multicolumn{2}{c}{6.79\%} \\
\bottomrule
\end{tabular}%
}
\end{table}

\begin{table}[]
\ra{1.3}
\caption{Time savings using dynamic HDL slicing}
\label{tab:timesaving}
\resizebox{\columnwidth}{!}{%
\begin{tabular}{lclcclccl}\toprule

\multicolumn{1}{c}{\textbf{}} & \multicolumn{2}{c}{Sandbox} &  & \multicolumn{2}{c}{Dhrystone} &  & \multicolumn{2}{c}{Coremark} \\ \cline{2-3} \cline{5-6} \cline{8-9}
\multicolumn{1}{c}{} & \multicolumn{2}{c}{Dynamic slicing time saving} &  & \multicolumn{2}{c}{Dynamic slicing time saving} &  & \multicolumn{2}{c}{Dynamic slicing time saving} \\ \hline
$inst\_dest\_bin$ & \multicolumn{2}{c}{16.91\%} &  & \multicolumn{2}{c}{5.53\%} &  & \multicolumn{2}{c}{3.33\%} \\
$inst\_src\_bin$ & \multicolumn{2}{c}{13.71\%} &  & \multicolumn{2}{c}{1.58\%} &  & \multicolumn{2}{c}{11.10\%} \\
\bottomrule
\end{tabular}%
}
\end{table}

\section{Conclusions}\label{conclusions}
Fault injection on RTL requires excessively long simulation time which prevents detailed reliability evaluation of hardware components with significant number of injections. This paper presents a method to speed-up fault injection campaigns by minimizing of fault injection locations. The method applies dynamic slicing on HDL to accurately pinpoint fault injection locations and allows injection of critical faults in these time windows. In this way, this paper narrows down the fault space and provides reduced simulation time. Moreover, average 5-10\% extra gain in simulation time for fault injection is a significant improvement of the total chip validation costs, as this phase is the most time consuming. The proposed method is language-agnostic and suitable for industrial grade EDA tool flows. Experimental results on industrial-size example show that we achieve significant speed-up of the fault injection simulation when comparing to the state-of-the-art flows.

\section*{Acknowledgement}

This research was supported by project RESCUE funded from the European Union's Horizon 2020 research and innovation programme under the Marie Sklodowaska-Curie grant agreement No 722325.

\bibliographystyle{IEEEtran}
\bibliography{ref}

\end{document}